\documentstyle[aps,preprint,eqsecnum,epsf,floats,amsmath]{revtex}
\begin{document}
\draft
\preprint{\vbox{\hbox{{\tt [ SOGANG-HEP 310/03 | gr-qc/0308052 ]}}}}
\title{Exactly soluble model for self-gravitating D-particles with the wormhole}
\author
{Won Tae Kim\footnote{electronic address:wtkim@mail.sogang.ac.kr} and
  Edwin J. Son\footnote{electronic address:sopp@string.sogang.ac.kr}}
\address{Department of Physics and Basic Science Research Institute,\\
         Sogang University, C.P.O. Box 1142, Seoul 100-611, Korea}
\date{\today}
\maketitle
\begin{abstract}
We consider D-particles coupled to the CGHS dilaton gravity and obtain
the exact wormhole geometry and trajectories of D-particles by
introducing the exotic matter. The initial static wormhole background
is not stable after infalling D-particles due to the classical
backreaction of the geometry so that the additional exotic matter
source should be introduced for the stability. Then, the traversable
wormhole geometry naturally appears and the D-particles can travel
through it safely. Finally, we discuss the dynamical evolution of the
wormhole throat and the massless limit of D-particles.
\end{abstract}

\section{Introduction}
Motivated by a funny fiction, Morris and Thorne \cite{mt} have studied
the possibility of wormholes and suggested some necessary properties
for traversable wormholes. They pointed out that in order to construct
a Lorentzian wormhole, the exotic matter is required, which is defined
as the matter violating the Weak Energy Condition (WEC) \cite{mt2}. It
has been defined by the matter violating the Null Energy Condition
(NEC) and it is claimed that its amount is related to the Averaged
Null Energy Condition (ANEC) \cite{v}.

As a toy model, based on the
Callan-Giddings-Harvey-Strominger(CGHS) dilaton gravity model
\cite{cghs}, exactly soluble two-dimensional wormholes solutions are
obtained \cite{hkl} by adding ghost fields which are nothing but the exotic
matter source.
On the other hand, the $N$-body self-gravitating system in the
Jackiw-Teitelboim (JT) model \cite{jt} which has the constant 
curvature scalar has been studied and the closed solutions are
obtained for the massive particles \cite{mro}.  This exactness without
approximations may be useful in studying some nonperturbative
physics. Recently, self-gravitating $N$-body motion described
by a slightly modified particle action was exactly solved for the
asymptotically flat spacetime by using the CGHS model \cite{ks}.

In this paper, as a natural extension, we would like to study an
interesting soluble wormhole model which is defined by D-particles
coupled to the CGHS dilaton gravity, where the kinetic term of the
usual scalar fields has a wrong sign to incorporate the exotic matter
in our starting action. In fact, the D-branes as nonperturbative
objects as RR charge carriers \cite{pol} have been
extensively studied in the string duality \cite{wit}. In the
present study, however, the D0-branes coupled to the gravity are
considered purely in order for the exact solubility of the geometry
and the trajectories of particles aiming to see how the particles as a
matter behave as time goes on and affect the geometry. The initial
static wormhole geometry is deformed by the infalling D-particles so
that the additional exotic source should be considered to maintain the
wormhole geometry at the latest time, which is in fact due to the
classical back reaction of the geometry after infalling D-particles.

In Sec.~\ref{sec:action}, we shall define our model as D-particles
coupled to the dilaton gravity described by the CGHS like model,
where the action for the D-particles is written by introducing einbeins.
Then, we shall show in Sec.~\ref{sec:op} that the wormhole will be
unstable after particles pass through it due to the back reaction of
the geometry. In order to maintain the wormhole structure, the
additional exotic source will be added to the original source in
Sec.~\ref{sec:maintain}, so the wormhole will be stabilized within
some constraints. Finally, summary and discussions are given in
Sec.~\ref{sec:discussion}.

\section{Action and energy-momentum tensors}
\label{sec:action}

We start with D-particles coupled to the two-dimensional dilaton gravity \cite{cghs}
with a conformal scalar ghost written by 
\begin{subequations}
\label{action}
\begin{eqnarray}
S &=& S_{DG} + S_g + S_D, \\
S_{DG} &=& \frac{1}{2\pi} \int d^2 x \sqrt{-g} e^{-2\phi} \left[ R +
  4(\nabla\phi)^2 + 4\lambda^2 \right], \label{action:gravity} \\
S_g &=& \frac{1}{2\pi} \int d^2 x \sqrt{-g} \left[ \frac{1}{2} \left(
    \nabla f \right)^2 \right], \label{action:ghost} \\
S_D &=& - \sum_{a=1}^N \int d^2 x \int d\tau_a \delta^2 ( x
  - z_a ( \tau_a ) ) m_a e^{-\phi (x)}  \sqrt{- g_{\mu\nu} (x)
  \frac{dz_a^\mu}{d\tau_a} \frac{dz_a^\nu}{d\tau_a}}, \label{action:particle}
\end{eqnarray}
\end{subequations}
where $g_{\mu\nu}$ and $\phi$ are the metric and the dilaton field,
and $\lambda^2$ is a cosmological constant. The scalar ghost $f$ with
the negative kinetic sign is necessary to construct wormholes. 
Note that $e_a$, $z_a$, and $m_a$ are the
einbeins, the particle coordinates, and mass for the $N$-particles,
respectively.
The Born-Infeld type action \cite{bi} for D-particles can be
written by an alternative form in terms of the
einbein variables,
\begin{equation}
S_D= \frac{1}{2} \sum_{a=1}^N \int d^2 x \int d\tau_a \delta^2 ( x
  - z_a ( \tau_a ) ) e_a ( \tau_a ) \left[ e_a^{-2} ( \tau_a )
  g_{\mu\nu} (x) \frac{dz_a^\mu}{d\tau_a} \frac{dz_a^\nu}{d\tau_a} -
  m_a^2 e^{-2\phi(x)} \right], \label{action:alpha}
\end{equation} 
for the massless limit in later.
Compared with the conventional massive particles, the mass term is
effectively interpreted as the dilaton-dependent mass term.  

From the action (\ref{action}) and (\ref{action:alpha}), 
the equations of motion for the metric, dilaton, ghost field, einbein,
and the coordinates are given by
\begin{subequations}
\label{eom}
\begin{eqnarray}
& & 2e^{-2\phi} \left[ \nabla_\mu \nabla_\nu \phi - g_{\mu\nu} \left(
  \Box \phi - (\nabla\phi)^2 + \lambda^2 \right) \right] =
  T_{\mu\nu}^g + T_{\mu\nu}^D, \label{em:gravity} \\
& & e^{-2\phi} \left[ R - 4(\nabla\phi)^2 + 4 \Box \phi + 4 \lambda^2
  \right] - \frac{\pi}{\sqrt{-g}} \sum_{a} \int d\tau_a \delta^2 ( x -
  z_a ) e_a m_a^2 e^{-2\phi} = 0, \label{em:dilaton} \\
& & \Box f = 0, \label{em:ghost} \\
& & e_a^{-2} g_{\mu\nu}(z_a) \frac{dz_a^\mu}{d\tau_a}
  \frac{dz_a^\nu}{d\tau_a} + m_a^2 e^{-2\phi(z_a)} = 0,
  \label{em:einbein}
\end{eqnarray}
and
\begin{equation}
g_{\mu\nu}(z_a) \frac{d}{d\tau_a} \left( e_a^{-1}
  \frac{dz_a^{\nu}}{d\tau_a} \right) + \Gamma_{\mu\alpha\beta}(z_a)
  e_a^{-1} \frac{dz_a^\alpha}{d\tau_a} \frac{dz_a^\beta}{d\tau_a} -
  m_a^2 e_a e^{-2\phi(z_a)} \frac{\partial\phi}{\partial z_a^\mu} =
  0, \label{em:geodesic}
\end{equation}
\end{subequations}
respectively, where the energy-momentum tensors due to the ghost field
and the point masses are written as
\begin{subequations}
\label{EMtensor}
\begin{eqnarray}
T_{\mu\nu}^g &=& - \frac{1}{2} \left( \nabla_\mu f \nabla_\nu f -
  \frac12 g_{\mu\nu} (\nabla f)^2 \right), \label{EMtensor:ghost} \\
T_{\mu\nu}^D &=& \frac{\pi}{\sqrt{-g}} \sum_a \int d\tau_a \delta^2
  (x-z_a) e_a^{-1} g_{\mu\alpha} g_{\nu\beta}
  \frac{dz_a^\alpha}{d\tau_a} \frac{dz_a^\beta}{d\tau_a},
  \label{EMtensor:ptl}
\end{eqnarray}
\end{subequations}
and einbein equation of motion (\ref{em:einbein}) was used to
eliminate the particle mass term. Then, 
combining Eqs.~(\ref{em:gravity}) and (\ref{em:dilaton}) yields
the following useful relation,
\begin{eqnarray}
e^{-2\phi} \left[ R + 2 \Box \phi \right] &=& \frac{\pi}{\sqrt{-g}}
  \sum_a \int d\tau_a \delta^2 (x-z_a) e_a \left[ e_a^{-2} g_{\mu\nu}
  \frac{dz_a^\mu}{d\tau_a} \frac{dz_a^\nu}{d\tau_a} + m_a^2 e^{-2\phi}
  \right] \nonumber \\ 
&=& 0, \label{eq:reduced}
\end{eqnarray}
which nicely vanishes by using the einbein equations of motion
(\ref{em:einbein}) and it is crucial to obtain the exact geometry and
particle trajectories without any approximations.

In the conformal gauge defined by $g_{+-} = - (1/2) e^{2\rho},
g_{--} = g_{++} = 0$, where $x^\pm = (x^0 \pm x^1)$, the above
equations of motion (\ref{eom}) are written as
\begin{eqnarray}
& & 2 e^{-2\phi} \left[ 2 \partial_+ \phi \partial_- \phi - \partial_+
  \partial_- \phi + \frac{1}{2} \lambda^2 e^{2\rho} \right] =
  T_{+-}^g + T_{+-}^D, \label{conf:geo} \\
& & 8 e^{-2(\rho+\phi)} \! \! \left[ \! \partial_+ \! \partial_- \!
  \rho \! + \! 2 \partial_+ \! \phi \partial_- \! \phi \! - \! 2
  \partial_+ \! \partial_- \! \phi \! + \! \frac{1}{2} \lambda^2
  e^{2\rho} \right] \! \! - \! 2 \pi e^{-2\rho} \! \sum_a \! \! \int \!
  \! d\tau_a \delta^2 \! (x \! - \! z_a \! ) e_a m_a^2 e^{-2\phi} \! =
  \! 0, \qquad \label{conf:dilaton} \\
& & e^{-2\rho} \partial_+ \partial_- f = 0, \label{conf:ghost} \\
& & e_a^{-2} e^{2\rho(z_a)} \frac{dz_a^+}{d\tau_a}
  \frac{dz_a^-}{d\tau_a} - m_a^2 e^{-2\phi(z_a)} = 0,
  \label{conf:einbein} \\
& & \frac{d}{d\tau_a} \left( e_a^{-1} \frac{dz_a^\pm}{d\tau_a} \right)
  + 2 e_a^{-1} \frac{\partial \rho(z_a)}{\partial z_a^\pm}
  \frac{dz_a^\pm}{d\tau_a} \frac{dz_a^\pm}{d\tau_a} + 2 m_a^2 e_a
  e^{-2(\rho + \phi)(z_a)} \frac{\partial \phi}{\partial z_a^\mp} = 0,
  \label{conf:geodesic}
\end{eqnarray}
with the constraint equations,
\begin{equation}
2 e^{-2\phi} \left[ \partial_\pm \partial_\pm \phi - 2 \partial_\pm
  \rho \partial_\pm \phi \right] = T_{\pm\pm}^g + T_{\pm\pm}^D,
  \label{conf:cons}
\end{equation}
and the energy-momentum tensors (\ref{EMtensor:ghost}) are
$T_{\pm\pm}^g = - (1/2) \partial_\pm f \partial_\pm f$ and
$T_{+-}^g = 0$. The source of D-particles is written as
\begin{subequations}
\label{conf:emt:ptl}
\begin{eqnarray}
T_{\pm\pm}^D &=& 2 \pi e^{-2\rho} \sum_a \int d\tau_a \delta^2 (x -
  z_a) e_a^{-1} \frac{e^{4\rho}}{4} \frac{dz_a^\mp}{d\tau_a}
  \frac{dz_a^\mp}{d\tau_a}, \label{conf:emt:p1} \\
T_{+-}^D &=& 2 \pi e^{-2\rho} \sum_a \int d\tau_a \delta^2 (x - z_a)
  e_a^{-1} \frac{e^{4\rho}}{4} \frac{dz_a^+}{d\tau_a}
  \frac{dz_a^-}{d\tau_a}. \label{conf:emt:p2}
\end{eqnarray}
\end{subequations}
Note that Eq.~(\ref{conf:emt:p2}) shows that the particle source is
not conformal while it vanishes with the help of
Eq.~(\ref{conf:einbein}) for the massless case. The key ingredient of
the exact solubility is due to Eq.~(\ref{eq:reduced}) written as in
the conformal gauge $\partial_+ \partial_- (\rho-\phi)=0$, and then
the residual symmetry can be fixed by choosing $\rho = \phi$ called 
Kruskal gauge.

For the sake of convenience, we now reparametrize as $m_a e_a d\tau_a =
d\lambda_a$ for the interesting massive case. The massless limit for
D-particles will be discussed in the final section briefly. From
Eqs.~(\ref{conf:einbein}) and (\ref{conf:geodesic}), after some
tedious calculations \cite{ks}, we get the following first order
differential equations,
\begin{equation}
\frac{dz_a^\pm}{d\lambda_a} = A_a^{(\pm)} e^{-2\rho(z_a)},
  \label{eq:particlem}
\end{equation}
where $A_a^{(\pm)}$ are integration constants which satisfy $A_a^{(+)}
A_a^{(-)} - 1 = 0$, and we choose $A_a^{(\pm)} > 0$ in order to make
$z_a^\pm$ be increasing functions with respect to
$\lambda_a$. Combining two equations of (\ref{eq:particlem}), we obtain
the trajectories of the particles,
\begin{equation}
z_a^+ = (A_a^{(+)})^2 \left( z_a^- + B_a \right), \label{rel:particle}
\end{equation}
where $B_a$ is an integration constant.
The trajectories of the particles are given as straight
lines in our coordinates similar to the massless source in the CGHS model.

Using Eqs.~(\ref{eq:particlem}) and (\ref{rel:particle}), the
energy-momentum tensors of the point particles
(\ref{conf:emt:ptl}) in the Kruskal gauge can be obtained as
\begin{subequations}
\label{EM:p}
\begin{eqnarray}
T_{++}^D &=& \frac{\pi}{2} \sum_a \frac{m_a}{A_a^3} \delta \left(
  \frac{x^+}{A_a^2} - x^- - B_a \right), \label{EM++} \\
T_{--}^D &=& \frac{\pi}{2} \sum_a m_a A_a \delta \left(
  \frac{x^+}{A_a^2} - x^- - B_a \right), \label{EM--} \\
T_{+-}^D &=& \frac{\pi}{2} \sum_a \frac{m_a}{A_a} \delta \left(
  \frac{x^+}{A_a^2} - x^- - B_a \right), \label{EM+-}
\end{eqnarray}
\end{subequations}
where we renamed $A_a^{(+)}$ to $A_a$ and $A_a^{(-)} = A_a^{-1}$.
At first sight, it seems for the energy-momentum tensors to vanish for
the massless limit; however, this is not the case since we have
already assumed the massive case when we reparametrize the proper
time. On the other hand, the solution of the ghost field is
$f=f_+(x^+)+f_-(x^-)$ from Eq.~(\ref{conf:ghost}).

\section{Wormhole and infalling D-particles}
\label{sec:op}

In this section, we shall obtain the wormhole geometry with the
infalling D-particles by assuming the exotic energy-momentum densities as
$T_{\pm\pm}^g = - \lambda^2$ $(f=\sqrt{2} \lambda (x^+-x^-))$, 
which is a special choice to get an
exact wormhole geometry. Then, integrating Eq.~(\ref{conf:geo})
with the energy-momentum tensor (\ref{EM+-}), we obtain the metric
solution,
\begin{equation}
e^{-2\rho} \! = \! a_+ \! ( x^+ \! ) \! + \! a_- \! ( x^- \! )
\! - \! \lambda^2 x^+ \! x^- \! - \! \frac{\pi}{2} \! \sum_a \! m_a
A_a \! \left( \! \frac{x^+}{A_a^2} \! - \! x^- \! - \! B_a \! \right)
\! \theta \! \! \left( \! \frac{x^+}{A_a^2} \! - \! x^- \! - \! B_a \!
\right),
\end{equation}
where $a_\pm(x^\pm)$ are integration functions determined by the
constraints (\ref{conf:cons}) as $\partial_\pm \partial_\pm a_\pm
(x^\pm) = \lambda^2$. Integrating it, the following metric solution is given by
\begin{equation}
e^{-2\rho} \! = \! D \! + \! C_+ x^+ \! + \! C_- x^- \! + \!
  \frac{1}{2} \lambda^2 \! \left( x^+ \! - \! x^- \right)^2 \! - \!
  \frac{\pi}{2} \! \sum_a \! m_a A_a \! \left( \! \frac{x^+}{A_a^2} \!
  - \! x^- \! - \! B_a \! \right) \! \theta \! \! \left( \!
  \frac{x^+}{A_a^2} \! - \! x^- \! - \! B_a \! \right),
\end{equation}
where $C_+$, $C_-$, and $D$ are integration constants.

\begin{figure}
\begin{center}
\leavevmode
\epsfxsize=0.5\textwidth
\epsfbox{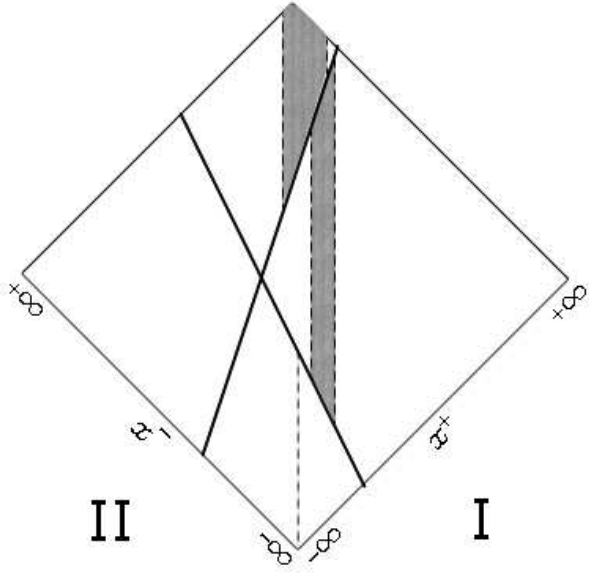}
\end{center}
\caption{This is a diagram of two-particle case; one starts from our
  universe (I), the other from the other universe(II). The solid lines
  denote the particle trajectories, the dotted lines are the horizons,
  and the shaded regions represent the future trapped regions.}
\label{fig:2-body}
\end{figure}

Next, we impose a boundary condition to make a wormhole 
at the region of $x^\pm \to - \infty$ where there are no infalling particles
by requiring the horizon coincidence condition as 
$\partial_+ e^{-2\rho} =\partial_-  e^{-2\rho} =0$.
The future and the past
horizons are explicitly obtained as
\begin{subequations}
\label{hor}
\begin{eqnarray}
0 &=& C_+ + \lambda^2 \left( x^+ - x^- \right) -
  \frac{\pi}{2} \sum_a \frac{m_a}{A_a} \theta \left( \frac{x^+}{A_a^2}
  - x^- - B_a \right), \\
0 &=& C_- - \lambda^2 \left( x^+ - x^- \right) +
  \frac{\pi}{2} \sum_a m_a A_a \theta \left( \frac{x^+}{A_a^2} - x^- -
  B_a \right),
\end{eqnarray}
\end{subequations}
respectively. So, the static wormhole condition
of $x^+ = x^-$ is applied to Eq.~(\ref{hor}) at the region of  $x^\pm \to -
\infty$, then we can fix $C_\pm$ and $D$,
\begin{equation}
C_+ = \frac{\pi}{2} \sum_a^{II} \frac{m_a}{A_a}, \quad
C_- = - \frac{\pi}{2} \sum_a^{II} m_a A_a, \textrm{ and }
D = \frac{M}{\lambda} - \frac{\pi}{2} \sum_a^{II} m_a A_a B_a,
\label{consts}
\end{equation}
where we labeled $\sum_a=\sum_a^I +\sum_a^{II}$, and 
$\sum_a^I = \sum_{a \in U_I}$ denotes the sum over all the
particles starting from our universe while $\sum_a^{II} = \sum_{a
\in U_{II}}$ means the sum over all the particles starting from the
other universe, {\it i.e.}, $U_I = \left\{ a | A_a^2 < 1 \textrm{ or }
  A_a^2 = 1, B_a > 0 \right\}$ and $U_{II} = \left\{ a | A_a^2 > 1
  \textrm{ or } A_a^2 = 1, B_a < 0 \right\}$. A diagram of two-body
case is shown in Fig.~\ref{fig:2-body}; one particle is included in
$U_I$ and the other is in $U_{II}$.
As a result, by using the identity, $1 - \theta (x) = \theta (-x)$,
the metric solution is represented as
\begin{eqnarray}
e^{-2\rho} &=& \frac{M}{\lambda} + \frac{1}{2} \lambda^2 \left( x^+ -
  x^- \right)^2 - \frac{\pi}{2} \sum_a^I m_a A_a \left(
  \frac{x^+}{A_a^2} - x^- - B_a \right) \theta \left(
  \frac{x^+}{A_a^2} - x^- - B_a \right) \nonumber \\
& & - \frac{\pi}{2} \sum_a^{II} m_a A_a \left( x^- - \frac{x^+}{A_a^2}
  + B_a \right) \theta \left( x^- - \frac{x^+}{A_a^2} + B_a \right),
  \label{sol:met}
\end{eqnarray}
which becomes static wormhole solution \cite{hkl}, $e^{-2\rho} = M /
\lambda + (1/2) \left( x^+ - x^- \right)^2$ for  $N = 0$.

Now, we obtain the horizon curves from the metric solution
(\ref{sol:met}),
\begin{subequations}
\label{horizon}
\begin{eqnarray}
0 &=& \lambda^2 \left( x^+ - x^- \right) - \frac{\pi}{2} \sum_a^I
  \frac{m_a}{A_a} \theta \left( \frac{x^+}{A_a^2} - x^- - B_a \right)
  + \frac{\pi}{2} \sum_a^{II} \frac{m_a}{A_a} \theta \left( x^- -
  \frac{x^+}{A_a^2} + B_a \right), \label{hor:future} \\
0 &=& \lambda^2 \left( x^+ \! - \! x^- \right) \! - \! \frac{\pi}{2}
  \sum_a^I m_a A_a \theta \! \left( \frac{x^+}{A_a^2} - x^- \! - B_a
  \right) \! + \! \frac{\pi}{2} \sum_a^{II} m_a A_a \theta \! \left(
  x^- \! - \frac{x^+}{A_a^2} + B_a \right). \label{hor:past}
\end{eqnarray}
\end{subequations}
The mass $(m_a)$ dependent terms eventually vanish at the asymptotic region of
$x^{\pm} \to -\infty$, and the past and future horizons are coincident
because of our boundary condition; however, both horizons are shifted
by the infalling particles in our model whereas only the future(past)
horizon is shifted for the lightlike infalling from universe I(II). It
can be easily seen from the horizon equation (\ref{horizon}) by simply
setting $N = 1$. As an illustration in Fig.~\ref{fig:2-body}, a
particle from our universe shifts both horizons to the right where the
future horizon is further shifted than the past horizon and the particle
from the other universe shifts both horizons to the left, in this
case, the past horizon is further shifted than the future
horizon. Therefore, the more particles pass through the wormhole, the
wider the gap between horizons is.

\section{Stability of the wormhole and D-particle trajectories}
\label{sec:maintain}

We have mentioned that the degenerate horizons in the far past are
split after particles passed through the wormhole. Thus, it 
is necessary to improve the exotic energy momentum tensor in order 
to maintain the wormhole structure even after all particles passed.
To do this, the original background exotic(ghost) energy is corrected
by ${\tilde T}_{\pm\pm}^g = T_{\pm\pm}^g + \Delta T_{\pm\pm}^g$ such
that the added term is $\Delta T_{\pm\pm}^g = \lambda^2  \beta_\pm
\left[ \theta \left( x^\pm - x_1^\pm \right) - \theta \left( x^\pm -
    x_0^\pm \right) \right]$ and $|\beta_\pm|<1$ are proportional
constants, and $x_0^\pm$ and $x_1^\pm$ satisfying $x_0^\pm < x_1^\pm$
are the coordinates where the additional field is turned on
and off. Then, a general metric solution is obtained as
\begin{eqnarray}
e^{-2\rho} \! &=& \! D \! + \! C_+ x^+ \! + \! C_- x^- \! + \!
  \frac{1}{2} \lambda^2 \! \left( x^+ \! - \! x^- \right)^2 \! - \!
  \frac{\pi}{2} \! \sum_a \! m_a A_a \! \left( \! \frac{x^+}{A_a^2} \!
  - \! x^- \! - \! B_a \! \right) \! \theta \! \! \left( \!
  \frac{x^+}{A_a^2} \! - \! x^- \! - \! B_a \! \right) \nonumber \\
& & + \frac{1}{2} \beta_+ \lambda^2 \left[ \left( x^+ - x_0^+
  \right)^2 \theta \left( x^+ - x_0^+ \right) - \left( x^+ - x_1^+
  \right)^2 \theta \left( x^+ - x_1^+ \right) \right] \nonumber \\
& & + \frac{1}{2} \beta_- \lambda^2 \left[ \left( x^- - x_0^-
  \right)^2 \theta \left( x^- - x_0^- \right) - \left( x^- - x_1^-
  \right)^2 \theta \left( x^- - x_1^- \right) \right],
\end{eqnarray}
where $C_+$, $C_-$, and $D$ are integration constants.
From the condition of $\partial_ + e^{-2\rho} =\partial_ -
e^{-2\rho}=0$, the future
and the past horizon are given as
\begin{subequations}
\label{hor2}
\begin{eqnarray}
0 &=& C_+ + \lambda^2 \left( x^+ - x^- \right) -
  \frac{\pi}{2} \sum_a \frac{m_a}{A_a} \theta \left( \frac{x^+}{A_a^2}
  - x^- - B_a \right) \nonumber \\
& & + \beta_+ \lambda^2 \left[ \left( x^+ - x_0^+ \right) \theta
  \left( x^+ - x_0^+ \right) - \left( x^+ - x_1^+ \right) \theta
  \left( x^+ - x_1^+ \right) \right], \\
0 &=& C_- - \lambda^2 \left( x^+ - x^- \right) +
  \frac{\pi}{2} \sum_a m_a A_a \theta \left( \frac{x^+}{A_a^2} - x^- -
  B_a \right) \nonumber \\
& & + \beta_- \lambda^2 \left[ \left( x^- - x_0^- \right) \theta
  \left( x^- - x_0^- \right) - \left( x^- - x_1^- \right) \theta
  \left( x^- - x_1^- \right) \right],
\end{eqnarray}
\end{subequations}
respectively.

Requiring the static wormhole boundary condition that the horizons (\ref{hor2})
be coincident along with the line $x^+ = x^-$ at $x^\pm
\to - \infty$, the same constants $C_\pm$ and $D$ with the previous
Eq.~(\ref{consts}) are derived.
Next, requiring the same boundary condition at the asymptotic region, $x^\pm \to +
\infty$ yields the following relations,
\begin{subequations}
\label{rel:beta}
\begin{eqnarray}
& & \beta_+ = \frac{\pi}{2 \lambda^2 \left( x_1^+ - x_0^+ \right)}
  \left( \sum_a^{I'} \frac{m_a}{A_a} - \sum_a^{II'} \frac{m_a}{A_a}
  \right), \\
& & \beta_-  = \frac{\pi}{2 \lambda^2 \left( x_1^- - x_0^- \right)}
  \left( - \sum_a^{I'} m_a A_a + \sum_a^{II'} m_a A_a \right),
  \label{rel:beta-}
\end{eqnarray}
\end{subequations}
where $\sum_a^{I'}$ denotes the sum over all non-static particles
starting from our universe and $\sum_a^{II'}$ denotes the sum over all
non-static particles starting from the other universe, {\it i.e.},
$U_{I'} = \left\{ a | A_a^2 < 1 \right\}$ and $U_{II'} = \left\{ a |
  A_a^2 > 1 \right\}$, where ``non-static'' means $dz_i^+ / dz_i^- =
A_i^2 \not = 1$. Note that the static cases represented by $A_a =1$
and $B_a <0$ or $B_a >0$ are canceled out in deriving these relations
from Eqs.~(\ref{hor2}), which means that only the traveling particles
contribute to the shift of the horizon, and consequently, the
correction to the exotic energy density is necessary for these
non-static infalling cases.
\begin{figure}
\begin{center}
\leavevmode
\epsfxsize=0.5\textwidth
\epsfbox{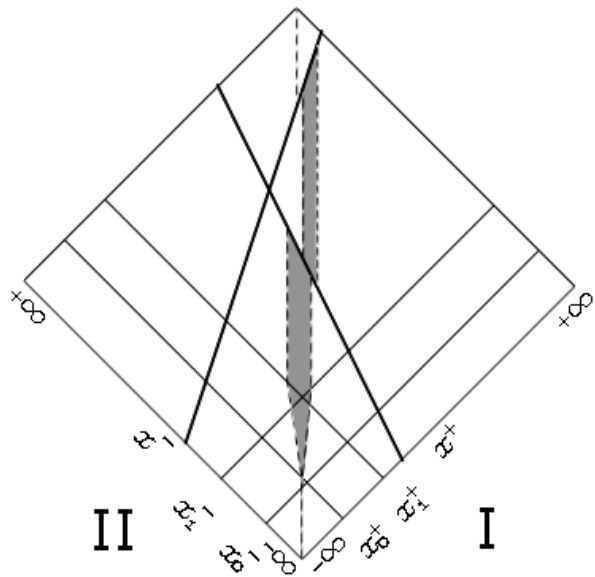}
\end{center}
\caption{This is a diagram of two-particle case; one starts from our
  universe (I), the other from the other universe(II), and the
  additional fields are turned on and off at $x_0^\pm$ and
  $x_1^\pm$. The thick solid lines denote the particle trajectories,
  the dotted lines show the horizons, and the shaded regions are the
  past trapped regions.}
\label{fig:maintain}
\end{figure}

For example, let us consider a single traveling particle$(N=1)$
starting from our universe, then Eqs.~(\ref{rel:beta}) read $\beta_+
>0$ and $\beta_- < 0$, which implies the left-handed(right-handed)
correction term $\Delta T_{++}^g$ $(\Delta T_{--}^g)$ should be
negative(positive) to recover the horizon shifts, and the additional
densities are explicitly $\Delta T_{++}^g= -(\pi m_1)/(2A_1(x_1^+
-x_0^+))$ and $\Delta T_{--}^g= +(\pi m_1 A_1)/(2(x_1^- -x_0^-))$ in
the finite interval $x^{\pm}_0 <x^\pm <x^{\pm}_1$. The left-handed
additional energy-momentum is larger than that of the right-handed one
for $x_1 ^+ =x_1^-$ and $x_0^+=x_0^-$ since the future horizon is
further right-shifted than the past horizon due to the traveling
particle starting from our universe as in Fig.~\ref{fig:2-body}. Note
that the right-handed source is positive since the past horizon was
shifted right.
 
As a result, the metric solution is given by
\begin{eqnarray}
e^{-2\rho} &=& \frac{M}{\lambda} + \frac{1}{2} \lambda^2 \left( x^+ -
  x^- \right)^2 - \frac{\pi}{2} \sum_a^I m_a A_a \left(
  \frac{x^+}{A_a^2} - x^- - B_a \right) \theta \left(
  \frac{x^+}{A_a^2} - x^- - B_a \right) \nonumber \\
& & - \frac{\pi}{2} \sum_a^{II} m_a A_a \left( x^- - \frac{x^+}{A_a^2}
  + B_a \right) \theta \left( x^- - \frac{x^+}{A_a^2} + B_a \right)
  \nonumber \\
& & + \frac{\pi \! \left( \sum_a^{I'} \! \! m_a \! - \!
  \sum_a^{II'} \! \! m_a  \right)}{4 \lambda^2 A_a \left( x_1^+ -
  x_0^+ \right)} \! \! \left[ \! \left( x^+ \! - \! x_0^+ \right)^2
  \theta \! \left( x^+ \! - \! x_0^+ \right) \! - \! \left( x^+ \! -
  \! x_1^+ \right)^2 \theta \! \left( x^+ \! - \! x_1^+ \right) \!
  \right] \nonumber \\
& & + \frac{\pi \! \left( \! - \! \sum_a^{I'} \! \! m_a A_a \! + \!
  \sum_a^{II'} \! \! m_a A_a \right)}{4 \lambda^2 \left( x_1^- - x_0^-
  \right)} \! \! \left[ \! \left( x^- \! - \! x_0^- \right)^2 \theta
  \! \left( x^- \! - \! x_0^- \right) \! - \! \left( x^- \! - \! x_1^-
  \right)^2 \theta \! \left( x^- \! - \! x_1^- \right) \! \right] \! ,
  \qquad \label{sol:geo}
\end{eqnarray}
which becomes again a stable wormhole solution, $e^{-2\rho} = M' /
\lambda + (1/2) \left( x^+ - x^- \right)^2$ at the region of $x^\pm
\to + \infty$, where $M' / \lambda = M / \lambda + (\pi / 2)
\sum_a^{I'} m_a A_a B_a - (\pi / 2) \sum_a^{II'} m_a A_a B_a 
-\left( \pi \left(x_0^+ + x_1^+ \right) / 4 \right) \left( \sum_a^{I'}
  m_a / A_a - \sum_a^{II'} m_a / A_a \right) - \left( \pi \left( x_0^-
    + x_1^- \right) / 4 \right) \left( - \sum_a^{I'} m_a A_a +
  \sum_a^{II'} m_a A_a \right)$ while it becomes $e^{-2\rho} = M /
\lambda + (1/2) \left( x^+ - x^- \right)^2$ at the asymptotic region
of $x^\pm \to - \infty$. Considering the one-particle case from our universe
($A_1^2 < 1$), the difference of $M$ between the latest time and the
initial time corresponding to the throat change is 
$\Delta M / \lambda = (\pi m_1/2) \left[ A_1 B_1 - ((x_0 +
  x_1)/2) A_1^{-1} (1-A_1^2) \right]$ for $x_0^+ = x_0^- = x_0$ and $x_1^+
= x_1^- = x_1$. The throat is unchanged, $\Delta M = 0$
especially for $(2 A_1^2 B_1)/(1-A_1^2) = x_0 + x_1$.

We now obtain the $i$-th particle trajectory in terms of the parameter
$\lambda_i$ by substituting Eq.~(\ref{sol:geo}) into
Eq.~(\ref{eq:particlem}),
\begin{equation}
\frac{dz_i^+}{d\lambda_i} = \left\{
\begin{array}{ll}
\sigma_i - \kappa_i z_i^+, & \textrm{ for } \gamma_i^2 = 0 \\
\frac{1}{2} \lambda^2 A_i \left[ \gamma_i^{-2} \left( \gamma_i^2 z_i^+
  - \zeta_i \right)^2 + \xi_i^2 \right], & \textrm{ for } \gamma_i^2
\not = 0
\end{array} \right. , \label{de:particle}
\end{equation}
where $\gamma_i^2 = \left[ (1-A_i^2)/A_i^2 \right]^2 + \beta_+ \theta
\left( z_i^+ - x_0^+ \right) - \beta_+ \theta \left( z_i^+ - x_1^+
\right) + \beta_- A_i^{-4} \theta \left( z_i^+ / A_i^2 - B_i - x_0^-
\right) - \beta_- A_i^{-4} \theta \left( z_i^+ / A_i^2 - B_i - x_1^-
\right)$, and $\sigma_i$, $\kappa_i$, $\zeta_i$, and $\xi_i^2$ are
defined by
\begin{eqnarray*}
\sigma_i &=& \frac{1}{2} \lambda^2 A_i \Bigg[ \frac{2M}{\lambda^3} +
  B_i^2 + \frac{\pi}{\lambda^2} \sum_a^I m_a A_a \left( B_a - B_i
  \right) \theta \left( z_i^+ \left( \frac{1}{A_a^2} - \frac{1}{A_i^2}
  \right) - \left( B_a - B_i \right) \right) \nonumber \\
& & + \frac{\pi}{\lambda^2} \sum_a^{II} m_a A_a \left( B_i - B_a
  \right) \theta \left( z_i^+ \left( \frac{1}{A_i^2} - \frac{1}{A_a^2}
  \right) - \left( B_i - B_a \right) \right) \nonumber \\
& & + \beta_+ \left( x_0^+ \right)^2 \theta \left( z_i^+ - x_0^+
  \right) - \beta_+ \left( x_1^+ \right)^2 \theta \left( z_i^+ -
  x_1^+ \right) \nonumber \\
& & + \beta_- \left( B_i + x_0^- \right)^2 \theta \left(
  \frac{z_i^+}{A_i^2} - B_i - x_0^- \right) - \beta_- \left( B_i +
  x_1^- \right)^2 \theta \left( \frac{z_i^+}{A_i^2} - B_i - x_1^-
  \right) \Bigg],
\end{eqnarray*}
\begin{eqnarray*}
\kappa_i &=& \lambda^2 A_i \Bigg[ \frac{\left( 1 - A_i^2 \right)
  B_i}{A_i^2} + \frac{\pi}{2\lambda^2} \sum_a^I m_a A_a \left(
  \frac{1}{A_a^2} - \frac{1}{A_i^2} \right) \theta \left( z_i^+ \left(
  \frac{1}{A_a^2} - \frac{1}{A_i^2} \right) - \left( B_a - B_i \right)
  \right) \nonumber \\
& & + \frac{\pi}{2\lambda^2} \sum_a^{II} m_a A_a \left(
  \frac{1}{A_i^2} - \frac{1}{A_a^2} \right) \theta \left( z_i^+ \left(
  \frac{1}{A_i^2} - \frac{1}{A_a^2} \right) - \left( B_i - B_a \right)
  \right) \nonumber \\
& & + \beta_+ x_0^+ \theta \left( z_i^+ - x_0^+ \right) - \beta_+
  x_1^+ \theta \left( z_i^+ - x_1^+ \right) \nonumber \\
& & + \beta_- \frac{B_i+x_0^-}{A_i^2} \theta \left(
  \frac{z_i^+}{A_i^2} - B_i - x_0^- \right) - \beta_-
  \frac{B_i+x_1^-}{A_i^2} \theta \left( \frac{z_i^+}{A_i^2} - B_i -
  x_1^- \right) \Bigg],
\end{eqnarray*}
$\zeta_i = \kappa_i / \lambda^2 A_i$, and $\xi_i^2 = 2 \sigma_i /
\lambda^2 A_i - \gamma_i^{-2} \zeta_i^2$.
Before getting solutions from the above two-differential
equations in Eq.~(\ref{de:particle}), we want to divide the worldline of
the $i$-th particle into some segments intersected by worldlines of
other particles. Then, $\gamma_i^2$, $\sigma_i$,
$\kappa_i$, $\zeta_i$ and $\xi_i^2$  
can be considered as constants in each segment and they are
represented by ${\gamma_i^{(r)}}^2$, $\sigma_i^{(r)}$, $\kappa_i^{(r)}$,
$\zeta_i^{(r)}$, and ${\xi_i^{(r)}}^2$, where $r$ is an index for the segments.
(i) For the case of ${\gamma_i^{(r)}}^2 = 0 = \kappa_i^{(r)}$, the solution
is easily obtained as $z_i^+ = \sigma_i^{(r)} \left( \lambda_i -
\Lambda_i^{(r)} \right)$, where $\Lambda_i^{(r)}$ is an integration
constant which can be fixed by the continuity with neighbor segments
and it is of relevance to choose the origin
of the parameter. The solution of $z_i^-$ is easily obtained from
Eq.~(\ref{rel:particle}). Note that this solution is adequate only if
$\sigma_i^{(r)} > 0$, because we assumed that $z_i$ is an increasing
function with respect to $\lambda_i$.
(ii) For the case of ${\gamma_i^{(r)}}^2 = 0$ and $\kappa_i^{(r)} > 0$,
substituting $\sigma_i^{(r)} - \kappa_i^{(r)} z_i^+ = {\eta_i^{(r)}}^{-1}$ into
Eq.~(\ref{de:particle}), we get $dz_i^+ / d\lambda_i =
{\kappa_i^{(r)}}^{-1} {\eta_i^{(r)}}^{-2} d\eta_i^{(r)} / d\lambda_i =
{\eta_i^{(r)}}^{-1}$. Thus, we obtain the particle motion,
$z_i^+ = \sigma_i^{(r)} / \kappa_i^{(r)} - {\kappa_i^{(r)}}^{-1} \exp \left[ -
  \kappa_i^{(r)} \left( \lambda_i - \Lambda_i^{(r)} \right)
\right]$. Note that this solution is valid  for $- \infty < z_i^+ <
\sigma_i^{(r)} / \kappa_i^{(r)}$.
(iii) For the case of ${\gamma_i^{(r)}}^2 = 0$ and $\kappa_i^{(r)} < 0$, the
solution is the same with the case of (ii), but the valid region is
different from that of (ii), $\sigma_i^{(r)} / \kappa_i^{(r)} < z_i^+
< + \infty$.
(iv) For the case of ${\gamma_i^{(r)}}^2 > 0$ ($\gamma_i^{(r)} > 0$) and
${\xi_i^{(r)}}^2 > 0$ ($\xi_i^{(r)} > 0$), substituting ${\gamma_i^{(r)}}^2 z_i^+ -
\zeta_i^{(r)} = \gamma_i^{(r)} \xi_i^{(r)} \tan \eta_i^{(r)}$ into
Eq.~(\ref{de:particle}), we get $z_i^+ = \zeta_i^{(r)} /
{\gamma_i^{(r)}}^2 + (\xi_i^{(r)} / \gamma_i^{(r)}) \tan \left[ (1/2) \lambda^2
  A_i \gamma_i^{(r)} \xi_i^{(r)} \left( \lambda_i - \Lambda_i^{(r)} \right)
\right]$ with the valid region, $-\infty < z_i^+ < +\infty$.
The other cases are similarly obtained, 
(v) for the case of ${\gamma_i^{(r)}}^2 > 0$ ($\gamma_i^{(r)} > 0$) and
${\xi_i^{(r)}}^2 = 0$ ($\xi_i^{(r)} > 0$), 
the geodesic equation yields $z_i^+ = \zeta_i^{(r)} / {\gamma_i^{(r)}}^2 - 2 /
\lambda^2 A_i {\gamma_i^{(r)}}^2 \left( \lambda_i - \Lambda_i^{(r)}
\right)$. Note that this solution is well-defined everywhere except at
the point $z_i^+ = \zeta_i^{(r)} / {\gamma_i^{(r)}}^2$.
(vi) For the case of ${\gamma_i^{(r)}}^2 > 0$ ($\gamma_i^{(r)} > 0$) and
${\xi_i^{(r)}}^2 = - \tilde {\xi_i^{(r)}}^2 < 0$ ($\tilde \xi_i^{(r)} > 0$),
the solution is given as $z_i^+ = \zeta_i^{(r)} / {\gamma_i^{(r)}}^2 + (\tilde
\xi_i^{(r)} / \gamma_i^{(r)}) \coth \left[ - (1/2) \lambda^2 A_i \gamma_i^{(r)}
  \tilde \xi_i^{(r)} \left( \lambda_i - \Lambda_i^{(r)} \right)
\right]$ with the valid ranges, $- \infty <
z_i^+ < \zeta_i^{(r)} / {\gamma_i^{(r)}}^2 - \tilde \xi_i^{(r)} /
\gamma_i^{(r)}$ and $\zeta_i^{(r)} / {\gamma_i^{(r)}}^2 + \tilde
\xi_i^{(r)} / \gamma_i^{(r)} < z_i^+ < + \infty$.
Finally, (vii) for the case of ${\gamma_i^{(r)}}^2 = - \tilde {\gamma_i^{(r)}}^2 < 0$
($\tilde \gamma_i^{(r)} > 0$) and ${\xi_i^{(r)}}^2 > 0$ ($\xi_i^{(r)} > 0$),
the solution is $z_i^+ = - \zeta_i^{(r)} / \tilde {\gamma_i^{(r)}}^2 +
(\xi_i^{(r)} / \tilde \gamma_i^{(r)}) \tanh \left[ (1/2) \lambda^2 A_i
  \tilde \gamma_i^{(r)} \xi_i^{(r)} \left( \lambda_i - \Lambda_i^{(r)} \right)
\right]$ with the valid range, $-
\zeta_i^{(r)} \tilde {\gamma_i^{(r)}}^2 - \xi_i^{(r)} / \tilde
\gamma_i^{(r)} < z_i^+ < - \zeta_i^{(r)} / \tilde {\gamma_i^{(r)}}^2 +
\xi_i^{(r)} / \tilde \gamma_i^{(r)}$.
\begin{figure}
\begin{center}
\leavevmode
\epsfxsize=0.5\textwidth
\epsfbox{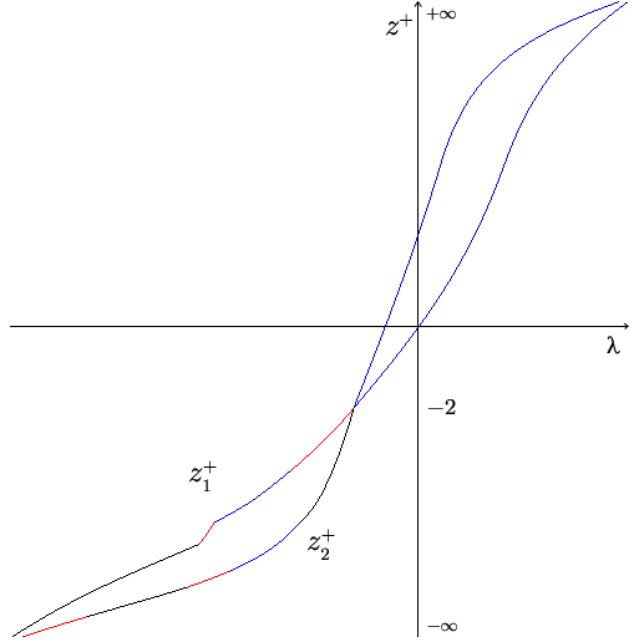}
\end{center}
\caption{Two particle trajectories which are continuous at each segment}
\label{fig:traj}
\end{figure}
As an illustration, we show the geodesic in Fig.~\ref{fig:traj} 
simply for the two-particle case by assuming the
constants as $m_1=m_2=M=\lambda=1$, $A_1^2=1/2$, $A_2^2=2$,
$B_1=-2$, $B_2=1$, $x_0^\pm=-10$, and $x_1^\pm=-5$. Then, the
proportional constants are $\beta_\pm = \sqrt{2} \pi / 20$, and $\gamma_1^2
= 1+(\sqrt{2} \pi / 20) \left[ \theta (z_1^+ +10) - \theta (z_1^+ +5)
  + 4 \theta (z_1^+ +6) - 4 \theta (z_1^+ +7/2) \right]$, $\gamma_2^2
= 1/4+(\sqrt{2} \pi / 20) \left[ \theta (z_2^+ +10) - \theta (z_2^+
  +5) + (1/4) \theta (z_2^+ +18) - (1/4) \theta (z_2^+ +8)
\right]$. The intersecting point between the two-particle worldlines is
$z^+ = -2$. Fig.~\ref{fig:traj} shows the particle trajectories with
respect to the parameter $\lambda$. The segments in the particle trajectories
are composed of the six segments in each trajectory.

\section{Discussion}
\label{sec:discussion}

We studied the soluble D-particle model coupled to the dilaton gravity
in 1+1 dimensions in terms of the well-known CGHS model, and
especially payed attention to the traversable wormhole construction and
its stability. The D-particles can travel from our universe to the
other universe with the help of the appropriately corrected exotic
matter such that the wormhole geometries 
appear at both asymptotic regions.

Generically, the coupling term
in front of the particle action (\ref{action:particle}) may be written as 
$e^{\alpha   \phi (x)}$, where $\alpha$ is a constant, 
then the conventional massive particle 
action is described for the case of $\alpha =0$, while the D-particle
case does for $\alpha =1$ as an open string coupling which has been 
our model. Unfortunately, for the former case, we still do not know
how to solve the model exactly; however, the latter case is exactly
solved and is still interesting in its own right and we hope it gives
some insights to the other physical models.

On the other hand, the wormhole throat is defined by the minimum
(radius) of $e^{-2\rho}$ in our model which is regarded as a radial
coordinate similar to the higher-dimensional analogue. One can easily
find the throat from $0 = \partial_1 e^{-2\rho(x^0,x^1)} = 4 \lambda^2
x^1 - (\pi/2) \sum_a^I m_a (A_a + A_a^{-1}) \theta ((A_a^{-2} - 1) x^0
+ (A_a^{-2} + 1) x^1 - B_a) + (\pi/2) \sum_a^{II} m_a (A_a + A_a^{-1})
\theta ((1 - A_a^{-2}) x^0 - (1 + A_a^{-2}) x^1 + B_a) + \beta_+
\lambda^2 [(x^0 + x^1 - x_0^+) \theta (x^0 + x^1 - x_0^+) - (x^0 + x^1
- x_1^+) \theta (x^0 + x^1 - x_1^+)] - \beta_- \lambda^2 [(x^0 - x^1 -
x_0^-) \theta (x^0 - x^1 - x_0^-) - (x^0 - x^1 - x_1^-) \theta (x^0 -
x^1 - x_1^-)]$. In a static wormhole state corresponding to $N=0$ with
$\beta_\pm = 0$, the throat radius is constant, which is seen from
$e^{-2\rho(x^1=0)}=M/\lambda$. However, in the dynamical unstable case
such as simply $N=1$ and $\beta_\pm = 0$, it is given by
$e^{-2\rho(x^1=x^1_{\mathrm{throat}})} = M/\lambda - (\pi/4) (3 x^0 +
25 \pi / 32 \lambda^2 ) \theta (3 x^0 + 25 \pi / 16 \lambda^2 )$,
where we put $A_1 = 1/2$, $B_1 = 0$ for simplicity, so that the throat
radius is suddenly grown up when the particle passes through it, then
it will eventually be shrunk to zero and the wormhole disappears. Thus
we required $\beta_\pm \not = 0$, in order to stabilize the wormhole
if $N \not = 0$.
 
Although it has not been explicitly shown in this paper, 
we have considered the
massless case with a reparametrization, $e_a d\tau_a
= d\lambda_a$. In this case, the particles become light-like
and conformal, which is easily seen by substituting $m_a = 0$ into
Eq.~(\ref{action:alpha}), and then they are independent of the dilaton
coupling. And this model can be also exactly solved, which gives
similar features. As we pointed out that the massive particles break
the residual conformal symmetry in the equation, $\partial_+ \partial_-
(\rho-\phi)=0$ and the model can not be solved in this
frame, however, in our D-particle model the symmetry is naturally
maintained by the dilaton coupling.
The energy-momentum tensors for the massless limit are
interestingly written as a delta-functional type, 
$T_{++}^D = (\pi/2) \sum_a^{I'} A_a \delta(x^+ - z_a^+)$, $T_{--}^D =
(\pi/2) \sum_a^{II'} A_a^{-1} \delta(x^- - z_a^-)$, and $T_{+-}^D =
0$, which appeared in the CGHS model.

The final comment to be mentioned is that the integration constants of
Eq.~(\ref{consts}) are determined by the infalling energy-momenta.
Note that we just introduced the constant $M$ in the constant $D$,
which seems to be of no relevance to the infalling energy. In fact, the
parameter $M$ characterizes the throat of the static wormhole since we
have assumed the static wormhole as an initial background geometry,
otherwise one should consider the collapsing normal matter which
produces the constant $M$ from the Minkowski spacetime. Therefore, it
should be related to the infalling energy to make the wormhole
geometry. In our analysis, since we introduced the static wormhole as
a background, the parameter just describes the throat of the
static wormhole before infalling D-particles.

\vspace{1cm}
{\bf Acknowledgments} \\
We would like to thank R.~Banerjee, J.H.~Cho, S.~Hyun, and P.~Oh
for exciting discussions. This work was supported by Korea Research
Foundation Grant (KRF-2002-042-C00010).

\end{document}